# Dynamic wavefront transformer based on a two-degree-of-freedom control system for 6-kHz mechanically actuated beam steering

**Authors:** Wenjun Deng[1,†], Weiming Zhu[1,†,*], Yuzhi Shi[2], Zhijun Liu[1], Guanxing Zang[1], Jin Qin[1] and Shiyu Zhu[1]

**Affiliations:**

[1]School of Optoelectronic Science and Engineering, University of Electronic Science and Technology, Chengdu, 610051, China

[2]National Key Laboratory of Science and Technology on Micro/Nano Fabrication, Department of Micro/Nano Electronics, Shanghai Jiao Tong University, Shanghai 200240, China

*Corresponding author: Weiming Zhu, zhuweiming@uestc.edu.cn

†Both authors contributed equally to this manuscript

**Abstract:** Vast tunable optical components are realized based on dynamic reconfigurations of the incident wavefronts, such as beam steering and tunable lens. However, the dominant paradigm of current wavefront reconfiguration technologies relies on complex control systems with degrees of freedom much larger than output wavefronts, e.g. beam steering based on spatial light modulator or phased array antennas. Here, we propose a new paradigm for dynamic reconfiguration of arbitrary output wavefronts using control systems with the same degrees of freedom. As an example, a wavefront transformer is demonstrated using an in-plane two-degree-of-freedom (2DOF) mechanical actuation system of metasurface doublet for semi-omnidirectional beam steering, which measured a 6-kHz modulation speed and a ±65.6° field of view. This paradigm can be applied to metasurface transformers for dynamic wavefront reconfiguration with any control system for vast applications, such as tunable lens, beam steering, and dynamic beam profiler, just to name a few.

# Dynamic wavefront transformer based on a two-degree-of-freedom control system for 6-kHz mechanically actuated beam steering

**Main Text:** Metasurfaces with spatial phase modulation down to subwavelength resolution show promising potential in passive wavefront conversion such as metalens [1, 2], hologram [3, 4], optical vortex beam [5, 6] and surface wave generation, etc [7, 8]. Although tunable and reconfigurable metasurfaces and metamaterials [9, 10] have been intensively demonstrated, the dynamic control of an arbitrary wavefront still follows the dominant paradigm widely used in spatial light modulators (SLM) or geometrical optics which cannot fully unleash the merits of metasurfaces in terms of tuning speed, flexibility, and compact size. For example, the pixel-by-pixel tuning method used in SLMs has now been intensively applied to reconfigurable metasurfaces [11-14]. Despite its complexity, the current control system, which has been developed for SLM and display devices for decades, has now reached the ceiling of integrated circuit (IC) fabrication and become the bottleneck of spatial resolution for wavefront reconfiguration [15]. As a result, the unit-cell by unit-cell tuning of visible and infrared metasurfaces is still a desired but unreachable technology. Metasurfaces for dynamic wavefront control are demonstrated either with lower resolution [16] or in lower frequency regions [17] to accommodate the state-of-the-art technology. Recently, metasurface lens doublets [18, 19] are demonstrated following the design route of geometrical optics, which unveil the potential of metasurface fast tuning devices based on mechanical actuation [20]. However, the flexibility and tuning speed of mechanically reconfigurable metasurfaces are limited due to the lacking of control freedom.

Dynamic wavefront control devices, such as tunable lens [21, 22], beam deflector [23, 24], and holographic display [25, 26] are longing for game-changing technologies with fast speeds, control flexibilities, design simplicities, and compact sizes. It is worth noting that the desired output wavefronts can be described by surface equations with finite degrees of freedom for most applications. Take beam steering as an example, the planar output wavefront can be described by a surface equation as $z=ax+by+c$, which has two degrees of freedom ($a, b$) as $c$ is uncorrelated to steering angles. On the other hand, the in-plane mechanical actuation system of metasurfaces also has two degrees of freedom with orthogonal displacement parameters along $x$- and $y$-directions ($S_x$ and $S_y$), which are enough for beam-steering wavefront control. More importantly, the arbitrary and abrupt phase change enabled by metasurfaces offers a concise approach for wavefront reconfiguration using a control system with the same degrees of freedom.

Here, we come up with a paradigm of dynamic wavefront transformers to control the output wavefront of an arbitrary incidence for tunable flat optics as shown in Fig. 1A. The incident wavefront is converted to a reconfigurable wavefront which can be described by surface functions with the same degrees of freedom as the output. The wavefront tuner maps the surface function of reconfigurable wavefront to output wavefront, which is dynamically controlled by a system of the same or higher degrees of freedom. A wavefront transformer with mechanical actuation is demonstrated as an example of this paradigm for beam steering functionalities as shown in Fig. 1B. The in-plane translation of a metasurface is a 2DOF system, which can be described by the displacement ($S_x$ and $S_y$) along two orthogonal directions. The wavefront transformer consists of two metasurfaces. The wavefront converter (metasurface A) is to convert an arbitrary incidence to a reconfigurable wavefront so that it can be tuned by the wavefront tuner (metasurface B) with in-plane translation while maintaining the output wavefront planar for beam steering purposes. Both metasurfaces A and B are in the *xy*-plane while the incident light is propagating along the *z*-direction. The metasurfaces, as shown in the insert of Fig. 1A, are composed of silicon rods patterned on top of a single crystal silicon wafer using deep reactive-ion etching (DRIE) process. Detailed fabrication parameters can be found in supplementary materials (section 1). The metasurfaces are designed for 10.6-μm incidence with abrupt phase change due to both the electromagnetic resonance and propagating feature of the silicon rods [27], as shown in Fig. 1B. The metasurfaces phase profiles are defined by the spatial arrangement of silicon rods with diameters $D$ ranging from 1.6 μm to 3.4 μm, which have a 5-μm hexagonal period and 6.5-μm in height. The phase shift of metasurface unit cells covers $2\pi$ with the transmittance above 60%, as shown in Fig. 1C.

The wavefront transformer is a 2DOF system with input control variables $(S_x, S_y)$ and output wave vector projections along $x$- and $y$-directions $(k_x, k_y)$, which is coupled by a matrix as shown in Fig. 2A. The derivation process of metasurface phase profiles and coupling matrix is shown in table S1. Here the metasurface A and B are positioned in the $z=-d$ and $z=0$ planes, respectively. For beam steering purposes, the spatial phase distribution of output wavefront $u_{Bt}$ can be described by a planar equation as

$$u_{Bt}(x,y,S_x,S_y)=a(S_x,S_y)x+b(S_x,S_y)y+c(S_x,S_y) \tag{1}$$

The planar phase distribution is achieved by linear superposition of incident phase distribution $u_{Bi}(x,y)$ of metasurface B and its abrupt phase change profile $\phi_B(x+S_x,y+S_y)$, which can be written as,

$$u_{Bt}(x,y,S_x,S_y)=u_{Bi}(x,y)+\phi_B(x+S_x,y+S_y) \tag{2}$$

where

$$u_{Bi}(x,y)=\frac{g_1}{2}x^2+\frac{g_2}{2}y^2$$

$$\phi_B(\alpha,\beta)=-\frac{g_1}{2}(x+S_x)^2-\frac{g_2}{2}(y+S_y)^2$$

As a result $k_x$ and $k_y$ are controlled by $S_x$ and $S_y$ by a diagonal matrix as follows,

$$\begin{cases} k_x(S_x,S_y)=-\frac{\partial u_{Bt}}{\partial x}=-a(S_x,S_y)=g_1 S_x \\ k_y(S_x,S_y)=-\frac{\partial u_{Bt}}{\partial y}=-b(S_x,S_y)=g_2 S_y \end{cases} \Rightarrow \begin{pmatrix} k_x \\ k_y \end{pmatrix}=\begin{pmatrix} g_1 & 0 \\ 0 & g_2 \end{pmatrix}\begin{pmatrix} S_x \\ S_y \end{pmatrix} \tag{3}$$

where $g_1$ and $g_2$ are constant.

The elevation angle $\theta$ and azimuth angle $\varphi$, as shown in the insert of Fig. 2A, are the functions of $S_x$ and $S_y$,

$$\theta(S_x,S_y)=\arcsin(\frac{k_s}{k_0}) \text{ and } \varphi(S_x,S_y)=\text{sign}(k_y)\cdot\arccos(\frac{k_x}{k_s})+180° \tag{4}$$

where $k_s=\sqrt{k_x^2+k_y^2}$.

Then the spatial phase distribution of the reconfigurable wavefront transmitted from metasurface A can be derived by $u_{Bi}$ using vector Rayleigh-Sommerfeld algorithm [28]. The abrupt phase change profile $\phi_A(x,y)$ of metasurface A can be obtained by,

$$\phi_A=u_{At}(x,y)-u_{Ai}(x,y) \tag{5}$$

where $u_{Ai}(x,y)$, $u_{At}(x,y)$ is the incident and output phase distribution of metasurface A.

Therefore, wavefront transformer can be designed by the derived $\phi_A$ and $\phi_B$ knowing the incident and output wavefront. More generalized and detailed derivation can be found in section 2 of the supplementary materials.

Fig. 2B shows the phase profiles of metasurface A (red solid line) and B (blue dash-dot line), when $g_1=g_2=4.234\times10^{-4}$ μm$^{-2}$ to maintain the rotation symmetry of the metasurfaces. The design parameters of metasurfaces A and B can be found in supplementary Fig. S1. Numerical analyses based on the finite element method (FEM) are performed to verify the beam steering function of the wavefront transformer. Fig. 2C and 2E show intensity distribution of 10.6-μm incidence in $xz$-plane when ($S_x=0, S_y=0$) and ($S_x=569$ μm, $S_y=0$), respectively. The simulation results show the steering angle of incident light can be controlled by translation of metasurface B, which is also verified by experimental results shown in Fig. 2D and 2F.

The beam steering function of the wavefront transformer is characterized using the experimental setup as shown in Fig. 3A. A 10.6-μm pulsed $CO_2$ laser (ACCESS LASER-AL30) with a full divergence angle of 5.5 mrad is used as the light source without collimation. The incident phase profile $u_{Ai}$ is estimated by using the ABCD law for Gaussian beam propagation [29]. To achieve a large actuation range, two linear motor stages (THORLAB-PT1-Z8) are used for x- and y-direction displacement of metasurface B. A beam profiler (WinCamD-FIR8-14-HR) is mounted on a circular track with a radius of 76 mm to measure the field of view. The fabrication results of metasurfaces are shown in Fig. 3B and 3C, which show the top view and cross-sections, respectively. Fig. 3D shows the output beam profile from the wavefront transformer when $S_x$ is varied from 0 μm to 1280 μm while $S_y=0$. The beam profiles expand along $x$-direction when the steering angle is increasing. This is mainly due to the projection of the beam on the detection plane, which results in a larger divergence angle with a smaller beam waist. The measured directivities normalized by the maximum point at 0° steering angle are shown in Fig. 3E, which indicates that the planar shape of the output wavefront is well maintained at different steering angles [30]. The measured field of view is ±65.6°, which is limited by the traveling range of the circular track. The 360° dynamic tuning of the azimuth angle is shown in supplementary materials (movie 1) while the elevation angle is limited by the size of the beam profiler sensor plane. The stationary spot in the detection plane is a damage point.

The angular velocities of beam steering can be expressed as (Supplementary section 2.4),

$$\begin{cases} \omega_\theta = \frac{1}{k_z k_s}(k_x \quad k_y)\begin{pmatrix} g_1 & 0 \\ 0 & g_1 \end{pmatrix}\begin{pmatrix} v_x \\ v_y \end{pmatrix} \\ \omega_\varphi = \frac{1}{k_s^2}(k_y \quad -k_x)\begin{pmatrix} g_1 & 0 \\ 0 & g_1 \end{pmatrix}\begin{pmatrix} v_x \\ v_y \end{pmatrix} \end{cases} \quad (6)$$

The modulation speed is predefined by the coupling matrix, i.e. $g_1$ in our case, and the mechanical actuation system. Fig. 4A shows measured elevation angle as a function of $S_r$ where $S_r=\sqrt{S_x^2+S_y^2}$. The experimental results agree with the simulation well, which shows the angular velocity of the elevation angle increases at a large steering angle when the linear actuation speed is a constant. For a given mechanical system, the modulation speed can be increased by choosing a large value of $g_1$. On the other hand, the azimuth angle only depends on the ratio of $S_x$ and $S_y$, as shown in Fig. 4B, which is not affected by $g_1$. Therefore, the tuning of elevation angle is chosen to demonstrate a fast beam steering with mechanical actuation by piezo actuators (THORLAB-PK4CMP1). The experimental setup is similar to that of the field of view measurement as shown in Fig. S2. The insert shows the transformer packaged with piezo actuators. The beam profiler is changed to a HgCdTe detector (TELEDYME-J15D16-M204-S01M-60) connected with an oscilloscope (TEKTRONIX-TDS-2024C) to capture the kHz beam steering process. A pinhole is placed right in front of the detector so that the change of elevation angle can be detected by the variation of light intensity. Fig. 4C shows the measured light intensity variation modulated by the beam steering process where the blue lines, dark lines, and red lines represent the measured light intensity, the modulation profile of beam steering, and the control signal respectively. The spikes of the measured light intensity are due to the pulsed output of the $CO_2$ laser. The modulation depth ($\frac{I_{max}-I_{min}}{I_{max}}$) remains above 50% with the modulation up to 6 kHz as shown in Fig. 4D. The modulation speed is capped by the resonance frequency (48.53 kHz) of the piezo actuation system loading with metasurface B, which is approximately 2 g in weight, as shown in Fig. S3. A fast modulation speed can be achieved by designing the metasurface B with a larger $g_1$ or reduce the weight of the metasurfaces by the backside thinning process of silicon wafer [31]. It should be pointed out that the wavefront transformer accommodates arbitrary incidence, which is essentially different from the beam steering devices based on refractive lens doublet limited by paraxial incidence and small steering angles.

The transformer design is not limited to beam steering functions. This paradigm can be applied to different applications with output wavefront which can be described by two decoupled variables, e.g. $u_{Bt}=ax^2+by^2$ or approximated by their Taylor's expansions (Supplementary materials section 2.5). Alvarez's lens is a special case for this paradigm, which has been widely demonstrated elsewhere. The modulation speed of demonstrated wavefront transformer is 6 kHz, which is limited by the bulky size of metasurface B. To our knowledge, it is the fastest mechanical beam steering demonstrated so far. The modulation speed of the wavefront transformer can be further pushed to 100 kHz using the piezo-actuation system or even MHz by microelectromechanical (MEMS) systems as discussed in Fig. S3 of the supplementary materials. It can be expected that vast applications of the wavefront transformer will be unlocked by adding new degrees of freedom to the mechanical actuation system, such as rotations and out-of-plane actuation

**Acknowledgments:**

　**Funding:**

　　National Natural Science Foundation of China grant 61975026

　　National Natural Science Foundation of China grant 61875030

　**Author contributions:**

　　Conceptualization: WZ

　　Methodology: WD, WZ, YS, JQ, SZ

　　Investigation: WZ, WD, YS, GZ

　　Visualization: WZ, WD, YS, GZ

　　Funding acquisition: WZ, ZL

　　Project administration: WZ, WD

　　Supervision: WZ

Writing – original draft: WD

Writing – review & editing: WZ, ZL

**Competing interests:** Authors declare that they have no competing interests.

**Data and materials availability:** All data are available in the main text or the supplementary materials.

**Supplementary Materials**

Materials and Methods

Supplementary Text

Figs. S1 to S3

Tables S1

References (32–36)

Movies S1

Dynamic wavefront transformer based on a two-degree-of-freedom control system for 6-kHz mechanically actuated beam steering

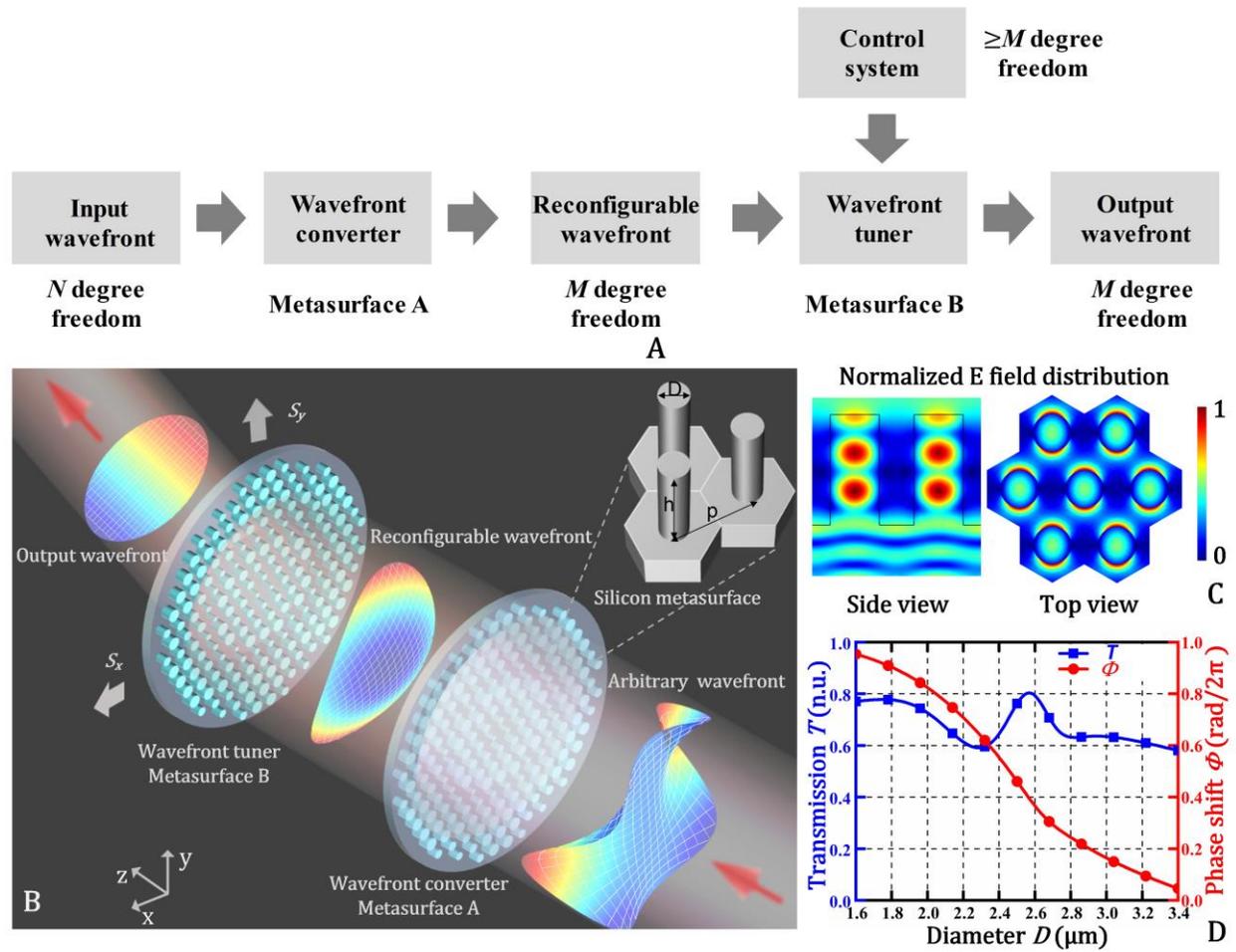

**Fig. 1. The concept of wavefront transformation based on reconfigurable metasurfaces.** (**A**) Schematic of reconfigurable metasurfaces for dynamic wavefront transformation. (**B**) Design of metasurface unit cells. (**C**) The transmission and phase shift of metasurface unit cells as a function of silicon rod diameters.

# Dynamic wavefront transformer based on a two-degree-of-freedom control system for 6-kHz mechanically actuated beam steering

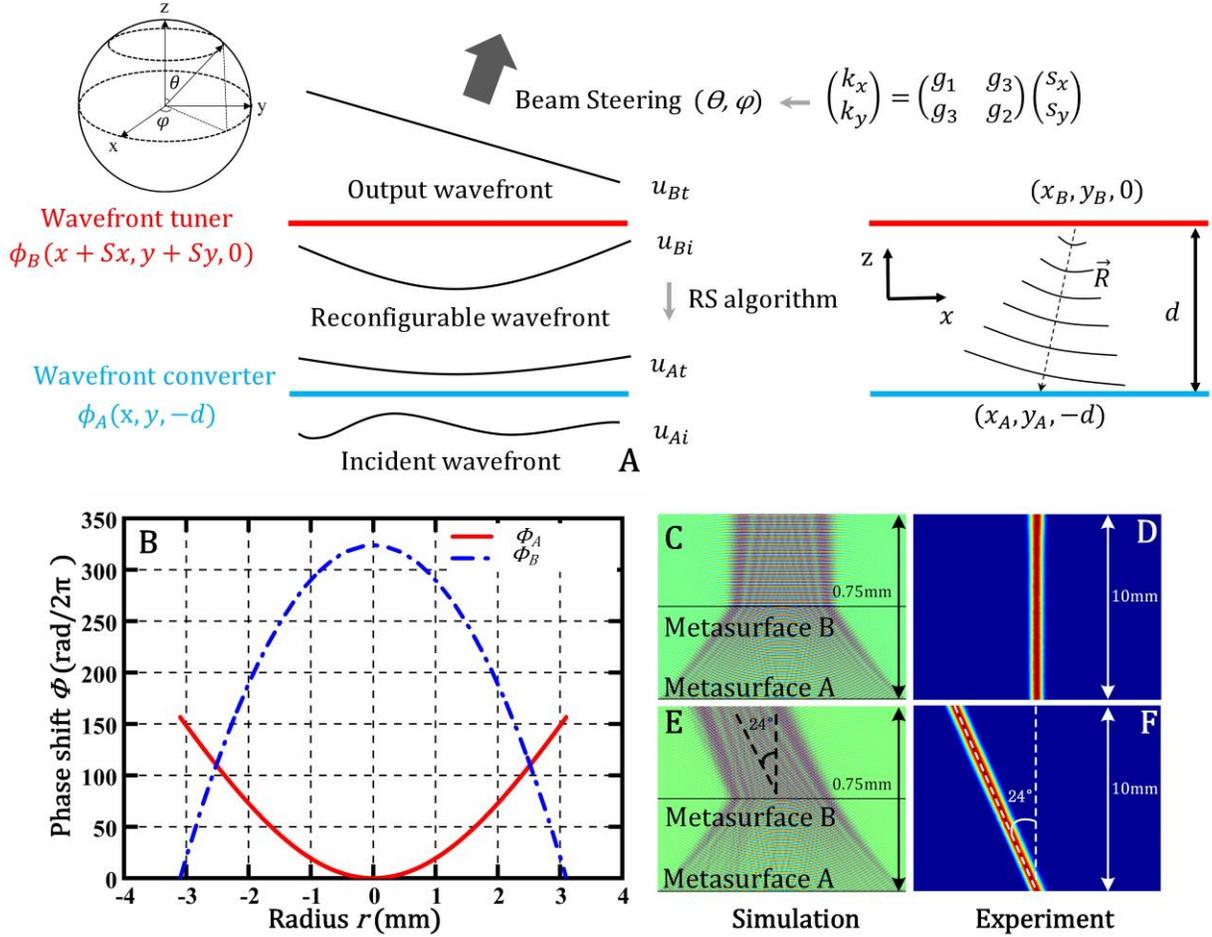

**Fig. 2. Theoretical and numerical analysis of wavefront transformation based on a two degree of freedom system.** (**A**) Spatial phase distribution of incident light wave is converted and reconfigured by wavefront converter (metasurface A) and tuner (metasurface B), respectively. The output wavefront is defined by two independent variables, which are the displacement of metasurface B at $x$- and $y$-directions ($S_x$ and $S_y$). The design of metasurface A and B are derived by using Rayleigh-Sommerfeld (RS) algorithm (**B**) Spatial distribution of phase shift for metasurface A (red solid line) and B (blue dash-dot line), respectively. The radius $r=\sqrt{x^2+y^2}$. (**C**), (**D**) (**E**) and (**F**) show simulation and experimental results when $S_x = S_y = 0$ μm (**C, D**) and $S_x = 569$ μm, $S_y = 0$ μm (**E, F**), respectively.

Dynamic wavefront transformer based on a two-degree-of-freedom control system for 6-kHz mechanically actuated beam steering

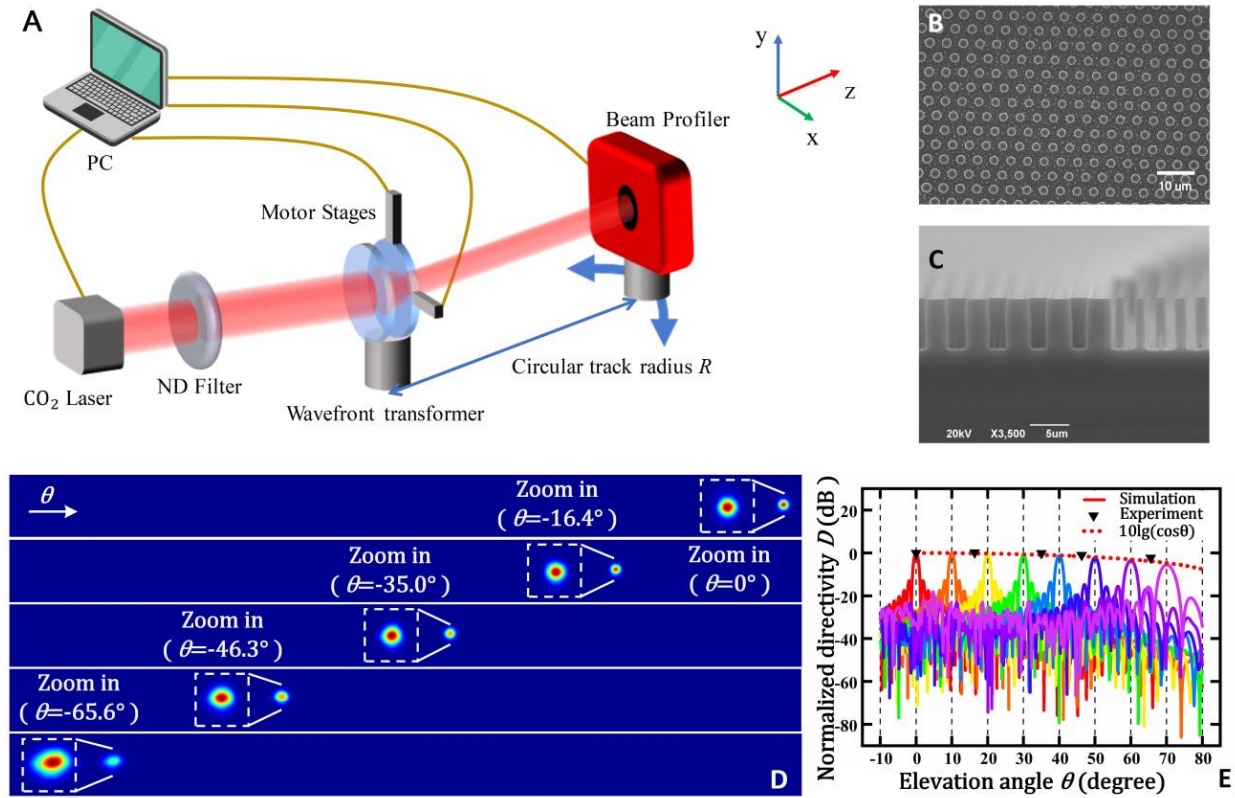

**Fig. 3. Experimental demonstration on beam steering based on reconfigurable metasurfaces.** (**A**) Experimental setup for beam profile and field of view characterization. The beam profiles are measured by a beam profiler mounted on a circular track with a radius of 76 mm to measure a large field of view. (**B**) and (**C**) are SEM graphs of the metasurface top view and cross-sections, respectively. (**D**) shows measured beam profile at different steering angles. (**E**) shows experimental (triangle symbol) and simulation results (solid lines) of normalized directivities at different steering angles.

Dynamic wavefront transformer based on a two-degree-of-freedom control system for 6-kHz mechanically actuated beam steering

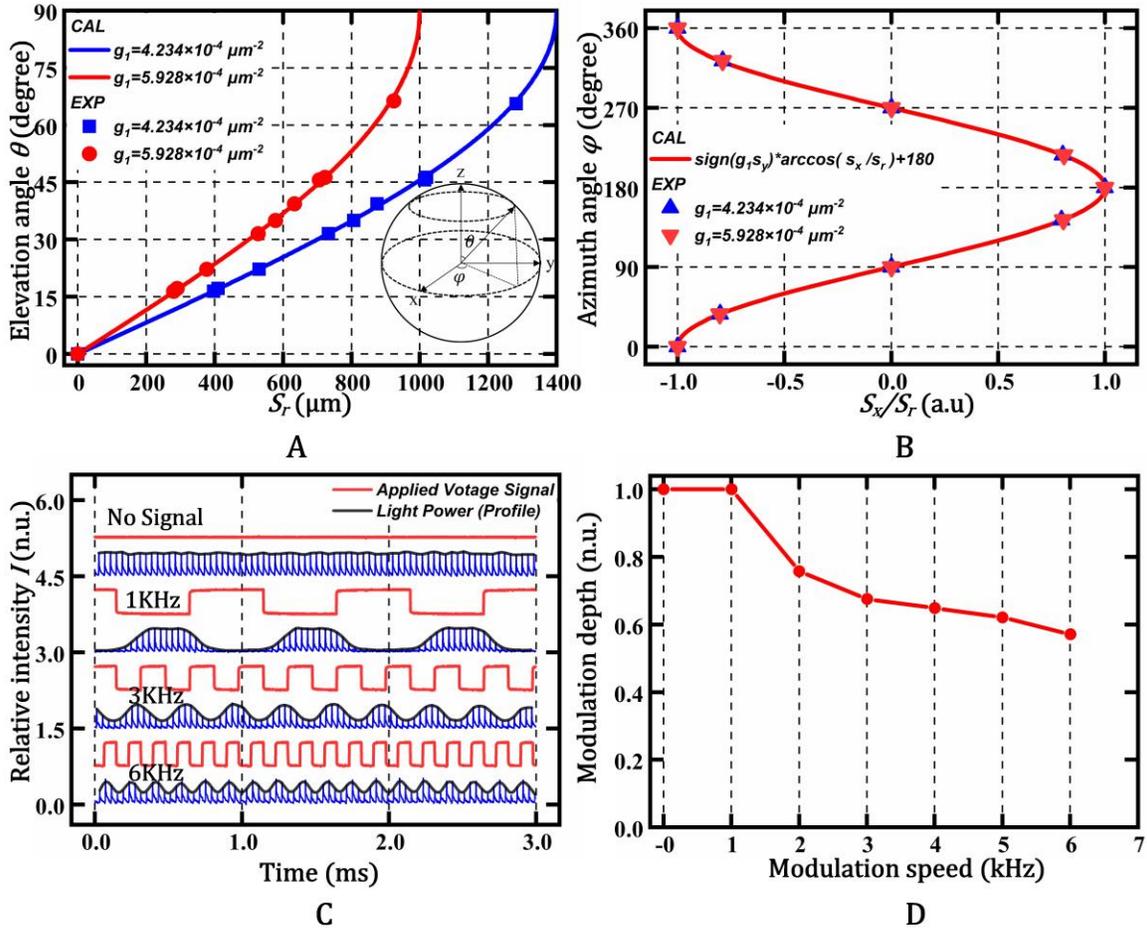

**Fig. 4. Experimental results on semi-omnidirectional beam steering and modulation speed.** (**A**) and (**B**) Elevation angle $\theta$ and Azimuth angle $\varphi$ as functions of displacements at different wavefront tuner designs, respectively. The solid lines show simulation results while the symbols represent experimental results. (**C**) shows dynamic responses of the beam steering devices at different actuation speeds. The spikes of the signals (blue lines) are due to the 25-kHz repetition rate of the pulse laser. The dark and red solid lines show the modulation profiles of the steering power and the control signal of the actuator, respectively. (**D**) The modulation depth ($\frac{I_{max}-I_{min}}{I_{max}}$) as a function of modulation speed.

# Supplementary Materials for

Dynamic wavefront transformer based on a two-degree-of-freedom control system for 6-kHz mechanically actuated beam steering

**Authors:** Wenjun Deng[1, †], Weiming Zhu[1, †, *], Yuzhi Shi[2], Zhijun Liu[1], Guanxing Zang[1], Jin Qin[1] and Shiyu Zhu[1]

Correspondence to: zhuweiming@uestc.edu.cn

**Materials and Methods**
1. Fabrication process
2. Derivation of metasurface phase profile and discussions
3. Discussion on nonplanar output wavefronts

Table S1
Figs S1 to S3

**Supplementary Text**

**1 Fabrication Processes**

The metasurfaces are fabricated by using a DRIE etching process of single crystal silicon wafer. First, the photo resist (AZ5214) is patterned on the top of silicon wafer followed by a 90-s pre-baking process with 95 °C. Then the metasurface structures are defined by 200-ms UV-exposure (NSR-2005i10D) with a photomask. The sample is post-baked for 90 s with the temperature of 100 °C after development process using JZX3038. Then the inductive coupled plasma (ICP) process is conducted by commercial STS DRIE system. Finally, the metasurface is realized by removing the residual photo resist.

**2 Derivation of metasurface phase profile and discussions**

The reconfigurable metasurfaces have two degrees of freedom $S_x$ and $S_y$, which are the input variables of wavefront transformer. The beam steering angle is defined by variables $a$ and $b$ of the output wavefront. Therefore, the derivation process of metasurface phase profile is to find a two-variables system to control $a$ and $b$ using $S_x$ and $S_y$ while maintaining the shape of output wavefront (for beam steering purpose $u_{Bt}$ should be planar). Here, the derivation process is as following. First, the output wavefront $u_{Bt}$ is used to derive $\phi_B$ and $u_{Bi}$. Here, $\phi_B$ is shifted by $S_x$ and $S_y$ while $u_{Bi}$ can be arbitrary form considering the wavefront conversion function of metasurface A. Then $u_{At}$ is derived by $u_{Bi}$ using vector Rayleigh-Sommerfeld algorithm based on the distance $d$ between the two metasurfaces, as shown in Fig. 2. Finally, the phase profile of metasurface A $\phi_A$ can be derived by $u_{At}$ and the incident phase profile $u_{Ai}$.

**2.1 Phase profile of metasurface B for beam steering**

For beam steering purpose, the output wavefront $u_{Bt}$ can be written as a planar form,

$$u_{Bt}(x,y,S_x,S_y)=a(S_x,S_y)x+b(S_x,S_y)y+c(S_x,S_y)=u_{Bi}(x,y)+\phi_B(\alpha,\beta) \quad (1)$$

where,

$$\begin{cases} \alpha = x + S_x \\ \beta = y + S_y \end{cases} \Rightarrow \begin{cases} \dfrac{\partial \alpha}{\partial x} = \dfrac{\partial \alpha}{S_x} = 1 \\ \dfrac{\partial \beta}{\partial x} = \dfrac{\partial \beta}{S_y} = 1 \end{cases}$$

(2)

According to (1) and (2),

$$\frac{\partial^2 u_{Bt}}{\partial x^2} = \frac{\partial^2 u_{Bi}}{\partial x^2} + \frac{\partial^2 \phi_B}{\partial \alpha^2} = 0$$

(3)

Therefore

$$\frac{\partial^2 u_{Bi}}{\partial x^2} = -\frac{\partial^2 \phi_B}{\partial \alpha^2}$$

(4)

Since $u_{Bi}$ is neither a function of $S_x$ nor $S_y$,

$$\begin{cases} \dfrac{\partial \dfrac{\partial^2 u_{Bt}}{\partial x^2}}{\partial S_x} = \dfrac{\partial \dfrac{\partial^2 u_{Bi}}{\partial x^2}}{\partial S_x} + \dfrac{\partial \dfrac{\partial^2 \phi_B}{\partial \alpha^2}}{\partial S_x} = \dfrac{\partial(\dfrac{\partial^2 \phi_B}{\partial \alpha^2})}{\partial \alpha} = 0 \\ \dfrac{\partial \dfrac{\partial^2 u_{Bt}}{\partial x^2}}{\partial S_y} = \dfrac{\partial \dfrac{\partial^2 u_{Bi}}{\partial x^2}}{\partial S_y} + \dfrac{\partial \dfrac{\partial^2 \phi_B}{\partial \alpha^2}}{\partial S_y} = \dfrac{\partial(\dfrac{\partial^2 \phi_B}{\partial \alpha^2})}{\partial \beta} = 0 \end{cases} \Rightarrow \dfrac{\partial^2 \phi_B(\alpha,\beta)}{\partial \alpha^2} \text{ is a constant}$$

(5)

According to (4) we can define

$$\frac{\partial^2 u_{Bi}}{\partial x^2} = -\frac{\partial^2 \phi_B}{\partial \alpha^2} = g_1$$

(6)

Similarly, we have

$$\frac{\partial^2 u_{Bi}}{\partial y^2} = -\frac{\partial^2 \phi_B}{\partial \beta^2} = g_2$$

(7)

where $g_1$ and $g_2$ are constant.

$$\frac{\partial^2 u_{Bt}}{\partial x \partial y} = \frac{\partial^2 u_{Bi}}{\partial x \partial y} + \frac{\partial \frac{\partial \phi_B}{\partial x}}{\partial y} = \frac{\partial^2 u_{Bi}}{\partial x \partial y} + \frac{\partial(\frac{\partial \phi_B}{\partial \alpha}\frac{\partial \alpha}{\partial x})}{\partial \beta}\frac{\partial \beta}{\partial y} = \frac{\partial^2 u_{Bi}}{\partial x \partial y} + \frac{\partial^2 \phi_B}{\partial \alpha \partial \beta} = 0$$

(8)

and

$$\begin{cases} \dfrac{\partial \dfrac{\partial^2 u_{Bt}}{\partial x \partial y}}{\partial S_x} = \dfrac{\partial \dfrac{\partial^2 u_{Bi}}{\partial x \partial y}}{\partial S_x} + \dfrac{\partial \dfrac{\partial^2 \phi_B}{\partial \alpha \partial \beta}}{\partial S_x} = \dfrac{\partial \dfrac{\partial^2 \phi_B}{\partial \alpha \partial \beta}}{\partial \alpha}\dfrac{\partial \alpha}{\partial S_x} = \dfrac{\partial^3 \phi_B}{\partial \alpha \partial \beta \partial \alpha} = 0 \\ \dfrac{\partial \dfrac{\partial^2 u_{Bt}}{\partial x \partial y}}{\partial S_y} = \dfrac{\partial \dfrac{\partial^2 u_{Bi}}{\partial x \partial y}}{\partial S_y} + \dfrac{\partial \dfrac{\partial^2 \phi_B}{\partial \alpha \partial \beta}}{\partial S_y} = \dfrac{\partial \dfrac{\partial^2 \phi_B}{\partial \alpha \partial \beta}}{\partial \beta}\dfrac{\partial \beta}{\partial S_y} = \dfrac{\partial^3 \phi_B}{\partial \alpha \partial \beta \partial \beta} = 0 \end{cases}$$

According to (8) and (9) we have

$$\frac{\partial^2 u_{Bi}}{\partial x \partial y} = -\frac{\partial^2 \phi_B}{\partial \alpha \partial \beta} = g_3 \tag{9}$$

$$\tag{10}$$

where $g_3$ is a constant.

According to (6), (7) and (10) $u_{Bi}$ and $\phi_B$ can be written as

$$\begin{cases} u_{Bi}(x,y) = \frac{g_1}{2}x^2 + \frac{g_2}{2}y^2 + g_3 xy + h_1 x + h_2 y + i_1 \\ \phi_B(\alpha,\beta) = -\frac{g_1}{2}\alpha^2 - \frac{g_2}{2}\beta^2 - g_3 \alpha\beta + h_3 \alpha + h_4 \beta + i_2 \end{cases}$$

$$\tag{11}$$

where $h_1, h_2, h_3, h_4, i_1, i_2$ are constants.

Therefore, the output wavefront can be tuned by changing $S_x$ and $S_y$ while remaining planar for beam steering purpose. The phase profile of metasurface B is defined by choosing the constants $g_1, g_2, g_3, h_1, h_2, h_3, h_4, i_1, i_2$, which have large design flexibility of metasurface B.

## 2.2 Phase profile of metasurface A

The spatial phase distribution of light wave $u_{At}(x,y)$ converted by metasurface A can be derived by $u_{Bi}$ using vector Rayleigh-Sommerfeld algorithm [32] as shown in Fig. 2,

$$U_{At}(x_A, y_A, -d) = -\frac{1}{2\pi} \iint U_{Bi}(x_B, y_B, 0) \frac{\partial G(\vec{R})}{\partial z_B}^* dx_B dy_B$$

$$\tag{12}$$

where,

$$U_{At}(x_A, y_A, -d) = A(x_A, y_A, -d) e^{-i u_{At}(x_A, y_A, -d)}$$
$$U_{Bi}(x_B, y_B, 0) = A(x_B, y_B, 0) e^{-i u_{Bi}(x_B, y_B, 0)}$$
$$R = \sqrt{(x_A - x_B)^2 + (y_A - y_B)^2 + d^2}$$
$$G(\vec{R}) = \frac{e^{ikR}}{R} \quad \frac{\partial G(\vec{R})}{\partial z_B}\bigg|_{z_B=0} = (ik - \frac{1}{R})\frac{e^{ikR}}{R}\frac{d}{R}$$

and $A(x,y,z)$ is the amplitude of the electrical field, which is also obtained by considering both the output wavefront and transmission of each unit cells. Here, the absorption of the air is not considered since $d$ is around 1 mm.

Then the phase profile of metasurface A is obtained by,

$$\phi_A = u_{At}(x_A, y_A, -d) - u_{Ai}(x_A, y_A, -d)$$

$$\tag{13}$$

where $u_{Ai}$ is the incident light wave phase distribution at metasurface A.

## 2.3 Discussion on beam steering angle

According to (1) and (11),

$$\begin{cases} a(S_x,S_y)=-(g_1S_x+g_3S_y+h_1+h_3) \\ b(S_x,S_y)=-(g_3S_x+g_2S_y+h_2+h_4) \\ c(S_x,S_y)=(-\frac{g_1}{2}S_x^2-\frac{g_2}{2}S_y^2-g_3S_xS_y+h_3S_x+h_4S_y+i_1+i_2) \end{cases}$$
(12)

Although $c(S_x,S_y)$ is a function of $S_x$ and $S_y$, it cannot affect the beam steering angle as it is irrelevant to $x$ and $y$. Let $h_1=h_2=h_3=h_4=0$ and based on Generalized Snell's Law,

$$\begin{cases} k_x(S_x,S_y)=-\frac{\partial u_{Bt}}{\partial x}=-a(S_x,S_y)=g_1S_x+g_3S_y \\ k_y(S_x,S_y)=-\frac{\partial u_{Bt}}{\partial y}=-b(S_x,S_y)=g_3S_x+g_2S_y \end{cases} \Rightarrow \begin{pmatrix} k_x \\ k_y \end{pmatrix} = \begin{pmatrix} g_1 & g_3 \\ g_3 & g_2 \end{pmatrix}\begin{pmatrix} S_x \\ S_y \end{pmatrix}$$
(13)

Here we define $k_0,k_x,k_y,k_z$ are the magnitude of the wave vector and its component along $x$, $y$ and $z$ directions, respectively. Equation (13) shows $S_x$ and $S_y$ can control the beam steering angle by changing $k_x$ and $k_y$ with a coupling matrix $\begin{pmatrix} g_1 & g_3 \\ g_3 & g_2 \end{pmatrix}$ The detailed expression of beam steering angles as functions of $S_x$ and $S_y$ are as following,

$$\theta(S_x,S_y)=\arcsin(\frac{k_S}{k_0})=\arcsin(\frac{\sqrt{(g_1^2+g_3^2)S_x^2+(g_2^2+g_3^2)S_y^2+2(g_1g_3+g_3g_2)S_xS_y}}{k_0})$$
(14)

$$\varphi(S_x,S_y)=sign(k_y)\cdot arccos(\frac{k_x}{k_S})+180°$$
$$=sign(g_3S_x+g_2S_y)\cdot arccos(\frac{g_1S_x+g_3S_y}{\sqrt{(g_1^2+g_3^2)S_x^2+(g_2^2+g_3^2)S_y^2+2(g_1g_3+g_3g_2)S_xS_y}})+180°$$
(15)

where $\theta$ is the elevation angle and $\varphi$ is the azimuth angle and,

$$k_s=\sqrt{k_x^2+k_y^2}$$
$$sign(x)=\begin{cases} 1, x\geq 0 \\ -1, x<0 \end{cases}$$

**2.4 Discussion on beam steering speed**

For a given mechanical actuation system, the tuning speed is limited by the intrinsic resonance frequency of the mechanical system since the actuation range decreases dramatically when the modulation frequency is beyond the intrinsic resonance [33]. However, the steering speed of a given mechanical system can be design by changing the coupling matrix $\begin{pmatrix} g_1 & g_3 \\ g_3 & g_2 \end{pmatrix}$.

Here, we define the actuation speed of metasurface B velocity as $v_x=(v_x,v_y)$, where $v_x=\frac{\partial S_x}{\partial t}$ and $v_y=\frac{\partial S_y}{\partial t}$, the angular velocity of beam steering $\omega_\theta$ and $\omega_\varphi$ are

$$\begin{cases} \omega_\theta = \dfrac{1}{k_z k_s}(k_x \;\; k_y)\begin{pmatrix} g_1 & g_3 \\ g_3 & g_2 \end{pmatrix}\begin{pmatrix} v_x \\ v_y \end{pmatrix} \\ \omega_\varphi = \dfrac{1}{k_s^2}(k_y \;\; -k_x)\begin{pmatrix} g_1 & g_3 \\ g_3 & g_2 \end{pmatrix}\begin{pmatrix} v_x \\ v_y \end{pmatrix} \end{cases}$$

(16)

In the metasurface design $g_1 = g_2 \neq 0$, $g_3 = 0$, and $S_r = \sqrt{S_x^2 + S_y^2}$, the beam steering angle and angular velocity can be simplified as,

$$\begin{cases} \theta(S_x, S_y) = arcsin\left(\dfrac{k_s}{k_0}\right) = arcsin\left(\dfrac{g_1 S_r}{k_0}\right) \\ \varphi(S_x, S_y) = sign(g_1 S_y) \cdot arccos\left(\dfrac{S_x}{S_r}\right) + 180° \\ \omega_\theta = \dfrac{g_1}{\sqrt{k_0^2 - g_1^2 S_r^2}} \vec{n_r} \vec{v_r} \\ \omega_\varphi = \dfrac{1}{S_r} |\vec{n_r} \times \vec{v_r}| \end{cases}$$

(17)

Where $\vec{n_r} = \dfrac{\vec{S_r}}{S_r} = (n_x, n_y)$

## 3 Discussion on nonplanar output wavefronts

In this section, we show that this paradigm can be applied to nonplanar output wavefronts with two degrees of freedom which can be described by a surface equation with two independent parameters.

### 3.1 Parabolic output wavefront

Here, the parabolic wavefront is chosen as an example.

$$u_{Bt}(x, y, S_x, S_y) = a(S_x, S_y)x^2 + b(S_x, S_y)y^2 + f(S_x, S_y)$$

(18)

Similar to the phase profile derivation process shown in section 2, the spatial phase distribution of metasurface B $\phi_B(\alpha, \beta)$ and its incidence $u_{Bi}(x,y)$ can be described by high order equations in $xy$ plane as,

$$\begin{cases} u_{Bi}(x,y) = \dfrac{f_1}{6}x^3 + \dfrac{f_2}{6}y^3 + \dfrac{f_3}{2}x^2 y + \dfrac{f_4}{2}xy^2 + \dfrac{g_1}{2}x^2 + \dfrac{g_2}{2}y^2 + g_3 xy + h_1 x + h_2 y + i_1 \\ \phi_B(\alpha, \beta) = -\left(\dfrac{f_1}{6}\alpha^3 + \dfrac{f_2}{6}\beta^3 + \dfrac{f_3}{2}\alpha^2 \beta + \dfrac{f_4}{2}\alpha\beta^2\right) + \dfrac{g_4}{2}\alpha^2 + \dfrac{g_5}{2}\beta^2 + g_6 \alpha\beta + h_3 \alpha + h_4 \beta + i_2 \end{cases}$$

(19)

where $f_1, f_2, f_3, f_4, g_1, g_2, g_3, g_4, g_5, g_6, h_1, h_2, h_3, h_4, i_1$ and $i_2$ are constants and

$$u_{Bt}(x, y, S_x, S_y) = u_{Bi}(x,y) + \phi_B(\alpha, \beta)$$

which can be written as,

$$u_{Bt}(S_x, S_y) = a(S_x, S_y)x^2 + b(S_x, S_y)y^2 + c(S_x, S_y)xy + d(S_x, S_y)x + e(S_x, S_y)y + f(S_x, S_y)$$

$$= \frac{g_1 + g_4 - f_1 S_x - f_3 S_y}{2} x^2 + \frac{g_2 + g_5 - f_2 S_y - f_4 S_x}{2} y^2 + (g_3 + g_6 - f_3 S_x - f_4 S_y)xy + (g_4 S_x + g_6 S_y + h_1 + h_3 - \frac{f_1 S_x^2 + f_4 S_y^2}{2} -$$

$$f_3 S_x S_y)x + (g_5 S_y + g_6 S_x + h_2 + h_4 - \frac{f_2 S_y^2 + f_3 S_x^2}{2} - f_4 S_x S_y)y + f(S_x, S_y)$$

(20)

Let $g_1 = g_2 = g_3 = g_4 = g_5 = g_6 = 0, h_1 = h_2 = h_3 = h_4 = 0$,

$$u_{Bt}(x,y,S_x,S_y) = -\frac{f_1 S_x + f_3 S_y}{2} x^2 -$$

$$\frac{f_2 S_y + f_4 S_x}{2} y^2 - (f_3 S_x + f_4 S_y)xy - (\frac{f_1 S_x^2 + f_4 S_y^2}{2} + f_3 S_x S_y)x - (\frac{f_2 S_y^2 + f_3 S_x^2}{2} + f_4 S_x S_y)y + f(S_x, S_y)$$

(21)

Then,

$$\begin{cases} a(S_x, S_y) = -\frac{f_1 S_x + f_3 S_y}{2} \\ b(S_x, S_y) = -\frac{f_2 S_y + f_4 S_x}{2} \\ c(S_x, S_y) = -(f_3 S_x + f_4 S_y) \\ d(S_x, S_y) = -(\frac{f_1 S_x^2 + f_4 S_y^2}{2} + f_3 S_x S_y) \\ e(S_x, S_y) = -(\frac{f_2 S_y^2 + f_3 S_x^2}{2} + f_4 S_x S_y) \end{cases}$$

(22)

Let $c(S_x, S_y) = 0$ to eliminate the cross term then,

$$\begin{cases} S_y/S_x = -f_3/f_4 \\ \text{or} \\ f_3 = f_4 = 0 \end{cases}$$

(23)

If $S_y/S_x = -f_3/f_4$ then $S_x$ and $S_y$ are coupled and the control system degenerates to one-degree-of-freedom system. Therefore, $f_3 = f_4 = 0$ and,

$$\begin{cases} u_{Bi}(x,y) = \frac{f_1}{6} x^3 + \frac{f_2}{6} y^3 + i_1 \\ \phi_B(\alpha, \beta) = -(\frac{f_1}{6} \alpha^3 + \frac{f_2}{6} \beta^3) + i_2 \end{cases}$$

(24)

$$u_{Bt}(x,y,S_x,S_y) = -\frac{f_1 S_x}{2} x^2 - \frac{f_2 S_y}{2} y^2 - \frac{f_1 S_x^2}{2} x - \frac{f_2 S_y^2}{2} y + f(S_x,S_y) = -\frac{f_1 S_x}{2}(x + \frac{S_x}{2})^2 - \frac{f_2 S_y}{2}(y + \frac{S_y}{2})^2 + f'(S_x,S_y)$$

(25)

with linear translation of the transformer along $x$- and $y$-directions of $-\frac{S_x}{2}$ and $-\frac{S_y}{2}$, respectively, the output wavefront can be written as,

$$u_{Bt}(x,y,S_x,S_y) = a(S_x,S_y)x^2 + b(S_x,S_y)y^2 + f'(S_x,S_y)$$

(26)

where,

$$\begin{cases} a(S_x,S_y) = -\dfrac{f_1 S_x}{2} \\ b(S_x,S_y) = -\dfrac{f_2 S_y}{2} \end{cases}$$

and $f'(S_x,S_y)$ is a function of neither $x$ nor $y$, which cannot change the output wavefront.

Therefore, the parabolic output wavefront can be controlled by $S_x$ and $S_y$. It should be pointed out that the linear translation of the transformer can be practically achieved by the mechanical actuation of both metasurface A and B with a collimated incidence.

### 3.2 Polynomial fitting of an arbitrary output wavefront by Taylor expansion

An arbitrary output wavefront can be described by the spatial phase distribution $u_{Bt}(x,y)$ at the plane of metasurface B (output plane). Here output wavefront for a hyperbolic lens is taken as an example to show how this paradigm works for wavefront transformer designs with functionalities other than beam steering. The spatial phase distribution of the output wavefront for a hyperbolic lens can be written as,

$$u_{Bt}(x,y) = (k_0 \sqrt{x^2+y^2+F^2} - F) \tag{27}$$

where $F$ is focal length of the hyperbolic lens and $k_0$ is wave vector of the incidence in free space. $u_{Bt}(x,y)$ can be rewritten by Taylor expansion near the origin of the output plane as,

$$u_{Bt}(x,y) = \sum_{n=1}^{\infty} \frac{\frac{1}{2}(\frac{1}{2}-1)\cdots(\frac{1}{2}-n+1)}{n!} \frac{k_0}{F} (x^2+y^2)^n \tag{28}$$

Therefore, by omitting high order terms with $n>1$, the approximate spatial phase distribution of a output wavefront for hyperbolic lens can be written as

$$u_{Bt}(x,y) = \frac{k_0}{2F}(x^2+y^2) = a(S_x,S_y)x^2 + b(S_x,S_y)y^2 \tag{29}$$

where $a(S_x,S_y) = b(S_x,S_y)$. This is a special case of Eq. (26), which are widely used in the designs of tunable lenses based on Alvarez principles. The focal length F is controlled by $S_x$ and $S_y$ as,

$$F = -\frac{k_0}{f_1 S_x} \tag{30}$$

where $S_x = S_y$.

As shown in Eq. (25), the transformer needs to be linear translated to keep the focal spots within the optical axis. Therefore, both metasurface A and B have to be controlled simultaneously for Alvarez lens, which is a well demonstrated example for this paradigm with a two degrees of freedom control system [34].

Dynamic wavefront transformer based on a two-degree-of-freedom control system for 6-kHz mechanically actuated beam steering

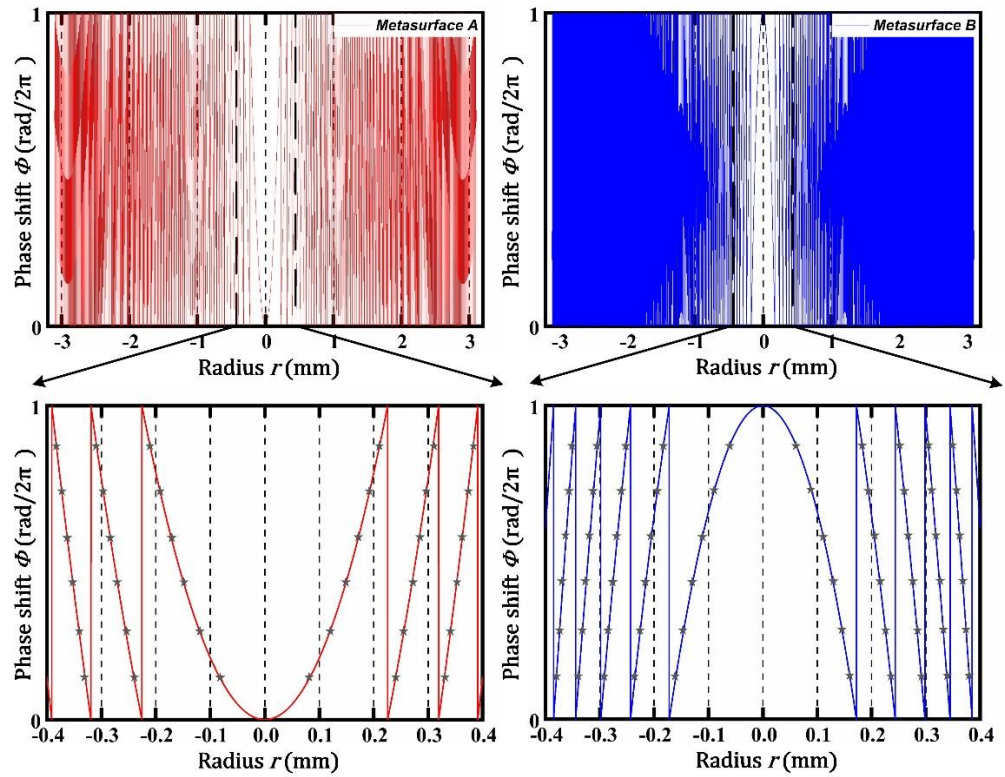

| Diameter (μm) | 3.40 | 2.98 | 2.69 | 2.54 | 2.41 | 2.25 | 2.06 | 1.74 |
| --- | --- | --- | --- | --- | --- | --- | --- | --- |
| Phase(rad) | 0 | $2\pi/8$ | $4\pi/8$ | $6\pi/8$ | $\pi$ | $10\pi/8$ | $12\pi/8$ | $14\pi/8$ |

**Fig. S1.**

Spatial phase distribution of metasurface A (left coloumn) and metasurface B (right coloumn). The second row shows the zoomed in views and the insert table shows the design parameters (diameter) of the silicon rods with 5-μm hexagonal period and 6.5-μm height. The symbols show the selected phase shifts and locations of the silicon rods.

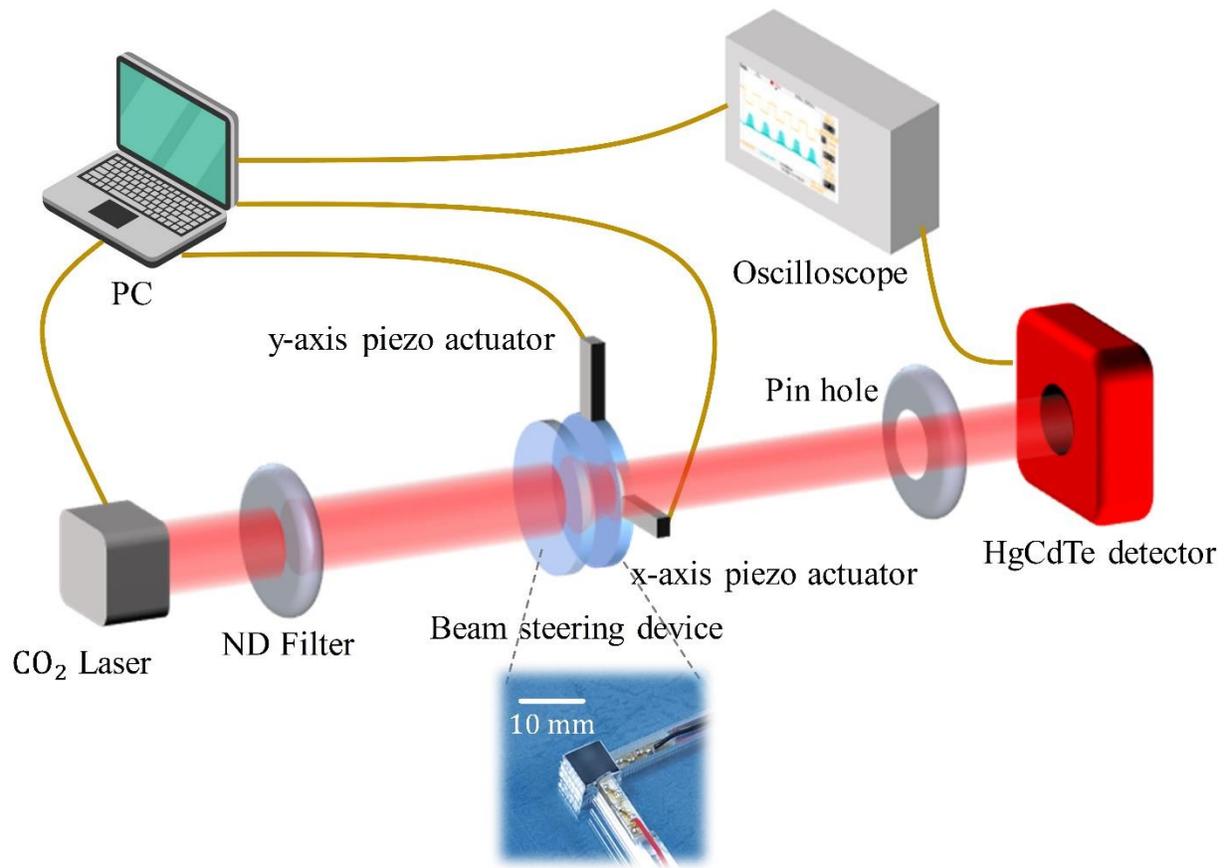

**Fig. S2.**
The sechmatic of experimental setup for beam steering speed characterization.

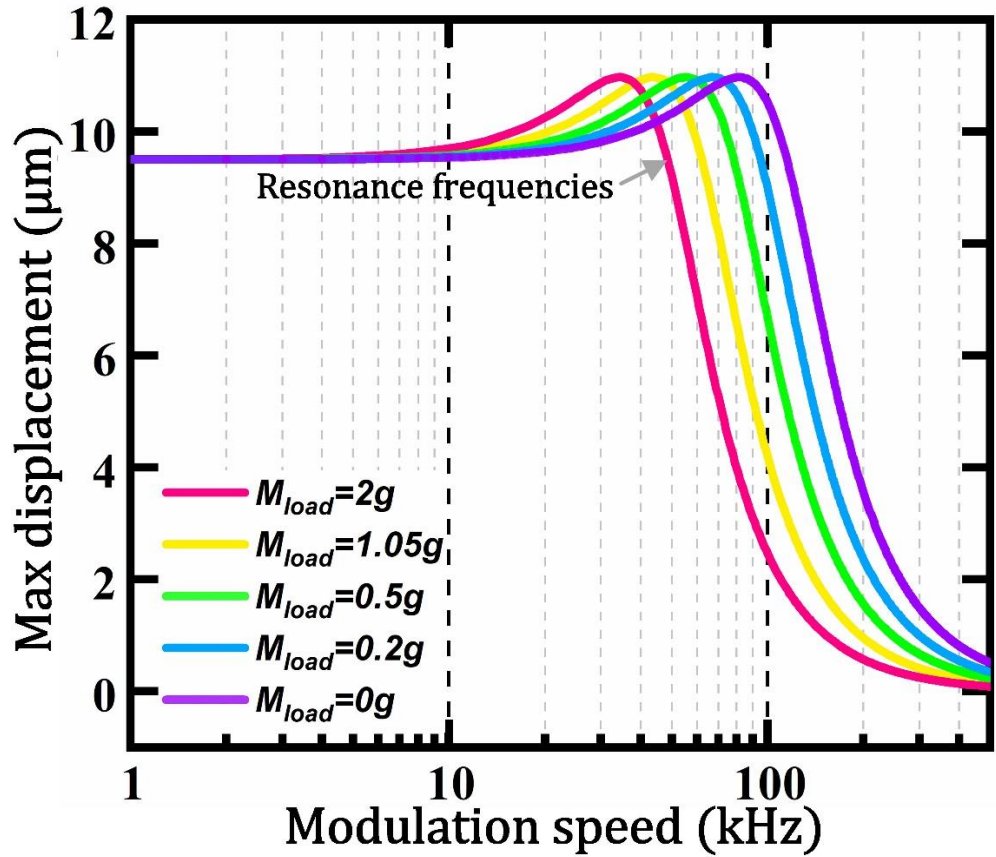

**Fig. S3.**

The maxiumium displacements as functions of modulation speed of the piezo-actuation mechanical system with different loadings [32]. The maxiumium displacements occurs when the modulation speed is close to the resoance frequencies of the mechanical system, which are senstive to the loading (mass of metasurface B). In our case, the modulation speed is measured to be 6 kHz and theoretically capped at 20 kHz due to the 2-g mass of the metasurfaces B. The thickness of the single crystal wafer is 1000 µm, which can be ten times thinner by using current wafer thinning technology [35]. Therefore, the modulation speed can approach 100 kHz using the pezo-actuation system. It can be expected that fast mechanical actuaction speed can be realized by MEMS/NEMS systems with much larger resonance frequencies [36].

**Table S1.**

The derivation process of metasurface phase profile.

| | Phase profiles | Expressions | |
|---|---|---|---|
| Output wavefront | $u_{Bt}$ | $a(s_x, s_y)x + b(s_x, s_y)y + c(s_x, s_y)$ | |
| Wavefront tuner | $\phi_B$ | $\phi_B(x + S_x, y + S_y, 0)$ | $u_{Bt} = u_{Bi} + \phi_B$ |
| Reconfigurable wavefront | $u_{Bi}$ | $u_{Bi}(x, y, 0)$ | RS algorithm |
| | $u_{At}$ | $u_{At}(x, y, -d)$ | |
| Wavefront converter | $\phi_A$ | $\phi_A(x, y, -d)$ | $\phi_A = u_{At} - u_{Ai}$ |
| Incident wavefront | $u_{Ai}$ | Aribitary form | |

Control variables $(S_x, S_y)$ ⇒ Output variables $(a, b)$

**Movie S1.**

Track of steering beam for 360° dynamic tuning of the azimuth angle while the elevation angle is limited by the size of the beam profiler sensor plane, where the stationary spot showing in the detection plane is a damage point.